\documentclass[aps,pre,reprint,twocolumn,groupedaddress,superscriptaddress,amsmath,amssymb,amsfonts,floats,floatfix,balancelastpage,showpacs]{revtex4-1}

\usepackage{epsfig} 
\usepackage{psfrag}
\usepackage{braket}
\usepackage{color}
\usepackage[usenames,dvipsnames]{xcolor}
\usepackage{textcomp}
\usepackage{graphicx}
\usepackage{subfigure}
\usepackage{multirow}
\usepackage{latexsym}
\usepackage{url}
\usepackage[english]{babel}
\usepackage[pdftex]{hyperref}

\usepackage{epstopdf}%This line makes .eps figures into .pdf - please comment out if not required.

\newcommand{\xv}{{\bf x}}

\newcommand{\uv}{{\bf u}}

\newcommand{\tv}{{\bf t}}

\newcommand{\grad}{{\bf \nabla}}
\newcommand{\zh}{\hat{z}}
\newcommand{\ph}{\hat{\phi}}

\begin{document}

\title{Chiral and achiral mechanisms of self-limiting, twisted bundle assembly}

\author{Gregory M. Grason}
\affiliation{Department of Polymer Science and Engineering, University of Massachusetts, Amherst, MA 01003, USA}

\begin{abstract}
A generalized theory of the self-limiting assembly of twisted bundles of filaments and columns is presented.  Bundles and fibers form in a broad variety of supramolecular systems, from biological to synthetic materials.  A widely-invoked mechanism to explain their finite diameter relies on chirality transfer from the molecular constituents to collective twist of the assembly, the effect of which frustrates the lateral assembly and can select equilibrium, finite diameters of bundles.  In this article, the thermodynamics of twisted-bundle assembly is analyzed to understand if chirality transfer is necessary for self-limitation, or instead, if spontaneously-twisting, achiral bundles also exhibit self-limited assembly.  A generalized description is invoked for the elastic costs imposed by twist for bundles of various states of intra-bundle order from nematic to crystalline, as well as a generic mechanism for generating twist, classified both by {\it chirality} but also the {\it twist susceptibility} of inter-filament alignment.  The theory provides a comprehensive set of predictions for the equilibrium twist and size of bundles as a function of surface energy as well as chirality, twist susceptibility, and elasticity of bundles.  Moreover, it shows that while spontaneous twist can lead to self-limitation, assembly of twisted {\it achiral} bundles can be distinguished qualitatively in terms of their range of equilibrium sizes and thermodynamic stability relative to bulk (untwisted) states.
\end{abstract}

\maketitle

\section{Introduction}

Bundles and fibers formed by supramolecular assembly are common architectures across a wide range of materials.  Fibers of extracellular proteins (e.g., cellulose, collagen, fibrin) consitute the basic structural and mechanical elements in plant and animal tissue~\cite{neville, fratzl2003}.  Beyond these functional architectures, the formation of fiber and cable assemblies of misfolded or mutant proteins are associated with various pathologies, from amyloidosis~\cite{chiti2006} to sickle cell anemia~\cite{mcdade1993}.   In synthetic systems, hierarchical assembly of 1D stacking constituents into multi-columnar bundles are widely observed in condensed phases of discotic liquid crystals~\cite{engelkamp1999, huang2013} and worm-like micelles~\cite{che2004}, organogels~\cite{douglas2009} and supramolecular ``polymers''~\cite{meijer01}.  

The functional (or pathological) properties of self-associated bundles and fibers, from their optical transmittance to their linear and non-linear mechanics, are highly dependent on their size distribution.  While most systems exhibit unlimited growth in the length of fibers, in many synthetic and biological assemblies, the lateral widths of assemblies are apparently well defined, or at least characterized by non-exponential distributions whose most probable size is finite and non-zero.  Motivated by the apparent reproducibility of this finite width, as well as its functional implications, a range of theoretical models have been proposed and explored to understand the finite fiber width as a result of {\it equilibrium} self-assembly.  The emergence of a finite-width falls outside of the canonical paradigms for equilibrium assembly~\cite{grason2016}, as generic considerations of surface energy in an aggregate imply that short range interactions typically favor macroscopically large dimensions (i.e. unlimited in size) in equilibrium.  As a result, physical mechanisms that have been invoked to explain finite bundle width, either resort to kinetically-arrested (i.e. nonequilibrium) aggregation models~\cite{wong2007}, or instead, to the presence of long-range interactions (i.e. much longer range than microscopic filament diameters)~\cite{dutta2016}.

One class of equilibrium mechanisms, which does not rely on explicitly long-range interactions , but nevertheless provides a thermodynamically consistent explanation for self-limitation of diameter is {\it chirality frustration}~\cite{grason2007, grason2009}.  Crudely speaking, this mechanism implies forces that, due to lack of mirror symmetry between constituents, favor local skews in the sub-unit packing~\cite{harris1999, kornyshev2007}.  When these local motifs propagate to larger length scales in the hierarchical bundle structure, they result in an intrinsic thermodynamic drive for collective twist.  Collective twist is incompatible with other types of order in the bundle (e.g. orientational, positional), and thus, gives rise to  elastic strains that build-up up with assembly size, ultimately providing an equilibrium mechanism to restrain the thermodynamic drive of surface energy towards unlimited sizes.

\begin{figure}
\centering
\includegraphics[width=1\linewidth]{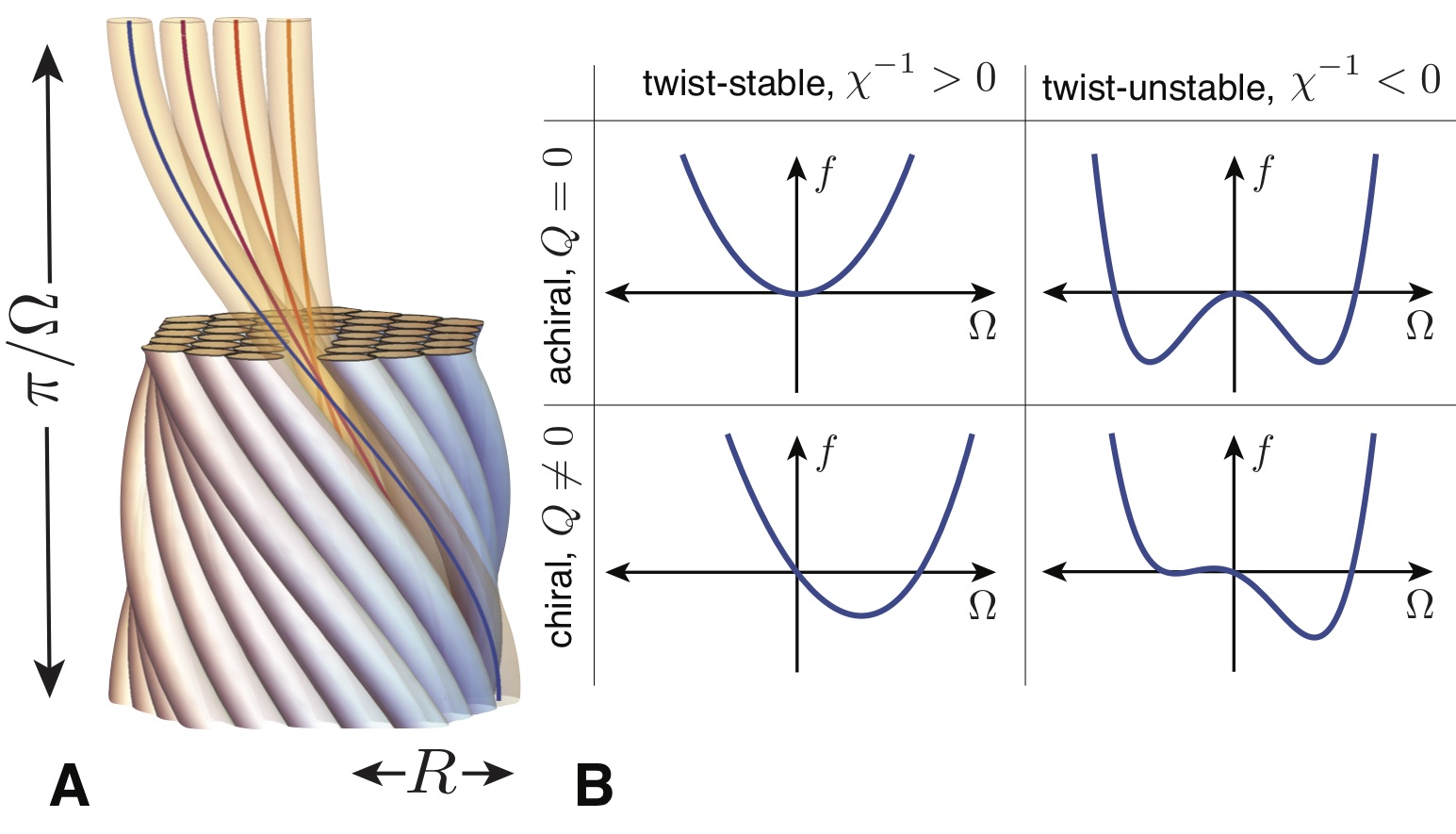}
\caption{In (A) a schematic of the geometry of a twisted bundle of filaments or columns.  (B) illustrates the distinct thermodynamics of inter-filament twist in bundles as a function of the {\it reduced chiralty} $Q$ and {\it reduced inverse twist susceptibility}, $\chi^{-1}$, which relate to the linear and quadratic coefficients of terms of the free energy density $f$ as function of bundle twist $\Omega$.  }
\label{introfig}
\end{figure}

Such a mechanism is attractive as an explanation of size selection in a range of biological bundle- and fiber-forming systems for two reasons.  First, the filaments themselves are generically composed from chiral building blocks (e.g. proteins, nucleic acids, polysaccharides).   Additionally, bundles or fibers of biofilaments are widely observed to exhibit collective, helical twist at the scale of the assembly~\cite{bouligand1985, bouligand2008}.  It is based on this reasoning that Makowski and Magdoff-Fairchild proposed a mechanism of self-limitation in twisted macrofibres of sickle-hemoglobin protofilaments~\cite{makowski1986}, a model that was quickly adapted to address the finite width of fibrin bundles~\cite{weisel1987}.  Subsequent to these pioneering studies, several distinct frameworks have been developed that integrate the chiral preference for collective bundle twist with various size-dependent elastic costs for assembly.  These include, the cost of intra-filament stretch in 3D solid bundles~\cite{turner2003} as proposed originally for the sickle hemoglobin and fibrin experiments, the cost of orientational gradients in {\it polymer nematic} bundles~\cite{rutenberg2014, rutenberg2018, murugesan2015}, the cost of variable inter-filament spacing in {\it 2D columnar} bundles~\cite{grason2007}, as well as the costs of inter-filament shearing in generalized models of {\it 3D crystalline} bundles~\cite{heussinger2011, grason2009,  addad2019}.   Along with these continuum models, coarse-grained simulation models of proto-filament assemblies with explicitly chiral interactions~\cite{yang2010}, all show well-defined regimes where the minimal free energy occurs for bundles at finite width, the size of which depends generically on elasticity of the assembly, cohesive forces driving assembly as well as ``strength'' of chirality.

While preferred-twist provides a thermodynamically consistent mechanism for self-limiting width of bundles of chiral systems, several recent studies of {\it achiral systems} raise the possibility that chirality at the building block scale may not be a necessary condition for self-limitation by twist.  For example, simulations of fibers formed by aggregation of ``sticky" semi-flexible chains~\cite{zierenberg2015, douglas2018} or by assembly supramolecular stacks of discotic molecules~\cite{wales2009, prybytak2012, dastan2017}, show the formation of spontaneously twisting structures without chiral building blocks.  The resulting double-twisted morphologies are superficially indistinguishable from chiral bundling systems, with the obvious exception that spontaneous twist is equally left- or right-handed in the achiral systems. Additionally, recent experimental studies of methylcellulose MC assemblies in aqueous solution have noted the emergence of fibrous aggregates whose radii are observed to be larger that the molecular thickness of a single MC strand~\cite{lott2013}, but consistently maintain a finite width of $\sim 18 \ {\rm nm}$ over a fairly broad range of assembly conditions.  It has recently been proposed that the finite width is consistent with a structural model of spontaneously, double twisting MC fiber morphology~\cite{schmidt2018, morozova2018}, which is loosely consistent with morphological observations, although a detailed determination of intra-fiber packing remains difficult to resolve.  As in the case of the achiral fiber simulations, the influence of molecular chirality in MC has initially been speculated to be weak~\cite{schmidt2018}, if it has any impact on the assembly at all.  

While it is perfectly understandable that twist emerges spontaneously in many achiral systems, and further may be a fairly generic effect in cohesive interactions between thread-like elements~\cite{cajamarca2014}, these observations raise important questions about the distinctions between mechanisms of achiral vs. chiral systems.  First, is it possible for spontaneous twist of an achiral fiber to give rise to thermodynamic self-limitation of assembly, or is intrinsic chirality essential to the finite size selection?  And second, if spontaneous twist does indeed lead to self-limitation, how is this {\it achiral} mechanism distinguishable from the {\it chiral} mechanism?  Specifically, how do the mechanisms differ in terms of thermodynamically selected sizes, pitches and regimes of stability of finite diameter fibers?

In this article, these questions are addressed in the context of continuum elasticity models of self-twisting filamentous bundles.  Symmetry considerations are used to construct the generic elastic costs of gradients in the orientational and positional order imposed by collective twist of bundles.  In particular, the thermodynamic drive for twist in bundles derives either from a {\it chiral} preference for inter-filament twist, or instead an {\it achiral} instability for spontaneous twist.  In this paper, this distinction is defined in terms of two parameters defined in detail in Sec.~\ref{sec: model} below:  $Q$ the reduced inverse pitch of preferred cholesteric twist; and $\chi^{-1}$, the reduced inverse twist susceptibility.  As illustrated in Fig. \ref{introfig}, these parameters are defined, respectively, by the first and second derivatives of the free energy with respect to bundle twist $\Omega$. The canonical case for chirality-driven twist corresponds to $Q \neq 0$ with a positive stiffness for twist, $\chi^{-1} \geq 0$.  A strictly achiral case of preferred spontaneous twist corresponds to $Q=0$ and negative inverse-susceptibility, $\chi^{-1} <0$.  In this article, I show that a sharp distinction between thermodynamics can be drawn between stable ($\chi^{-1} >0$) and unstable ($\chi^{-1}<0$) twist thermodynamics.  More specifically, the full range of thermodynamic behavior of self-twisting bundles is controlled by a single (dimensionless) combination of inverse pitch and susceptibility, $Q^2 \chi^3$.  This distinction can ultimately be traced to the variation of equilibrium bundle twist with size and the mechanical costs of intra-bundle deformation:  bending and shears of 2D and/or 3D inter-filament order.  The equilibrium variation of twist with bundle size can be directly related to the equilibrium size of bundles, and its dependence on the surface energy parameterizing inter-filament cohesion of the structure.  A central finding of this study is the sharp distinction between twist-stable and twist-unstable bundles with regard to the maximum range self-limited diameter of bundles and the critical value of surface energy that separates self-limited (twisted) and bulk (untwisted) assembly.  While the maximum self-limiting size of twist-stable ($\chi^{-1}>0$) bundles can grow arbitrarily large with vanishing chirality, the range of thermodynamic stability of such finite bundles also vanishes with $Q\to 0$.  In contrast, while the self-limiting sizes of twist-unstable ($\chi^{-1} <0$) bundles are largely independent of chirality, remaining limited ``microscopic" dimensions, their self-limited state is more stable relative to bulk assembly, and remains so, even as $Q\to0$.  

The remainder of this article is organized as follows.  In Sec. \ref{sec: model}, I briefly introduce the continuum model for self-twisting, cohesive bundles possessing various states of liquid-crystalline and crystalline order.  Then, in Sec.~\ref{sec: thermo} I describe the thermodynamics of equilibrium twist and size of bundles as functions of elastic parameters, surface energy and the driving forces for twist, first for 2D columnar order, and then 3D solid order.  In Sec. \ref{sec: discuss}, I discuss the implications of the distinctions between twist-stable and twist-unstable bundles for studies of chiral and achiral fiber assembly, and further, outline some open questions regarding the connections between microscopic descriptions of inter-molecular forces in bundles and continuum parameters describing the mesoscale behavior.

\begin{figure*}
\centering
\includegraphics[width=0.7\linewidth]{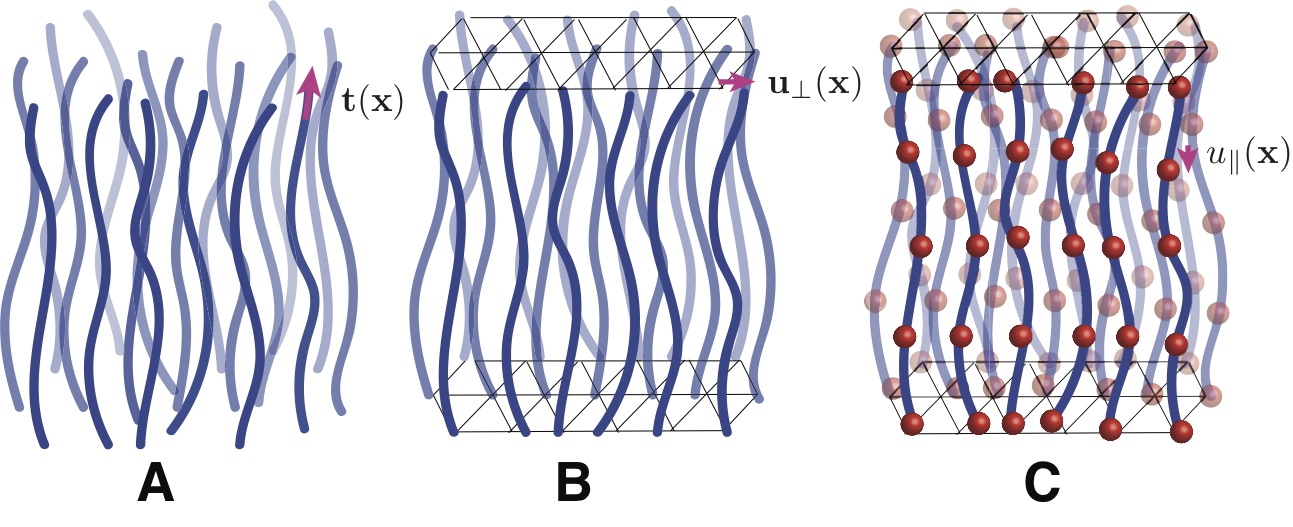}
\caption{Cartoons highlight the distinct types of order in bundle assemblies, and their associated order parameters:  (A) orientational order of backbones in nematically ordered bundles, with director $\tv(\xv)$ describing distortions; (B) transverse 2D  order inter-filament positions in columnar ordered bundles, with $\uv_\perp (\xv)$ indicated the transverse displacement relative to 2D lattice positions; and (C) 3D solid order of crystalline bundles, in which mass points (shown as red spheres) maintain long-range axial correlations between filaments, and with deflections from prefect ``layered order" described by longitudinal displacement $u_\parallel (\xv)$.  }
\label{orderfig}
\end{figure*}

\section{Generalized elasticity model of self-twisting, cohesive bundles}

\label{sec: model}

Here I introduce the model of helically-twisted bundles of columns and filaments.  Bundles are assumed to follow the ``double-twist" geometry,  familiar as local motif of liquid crystal blue phases~\cite{wright_mermin1989}, 
\begin{equation}
\label{eq: tv}
\tv(\xv) \simeq \zh + \Omega \rho \ph ,
\end{equation}
where $\tv(\xv)$ is the backbone orientation of constituent fibers, $\zh$ is the mean twist axis of the bundle, $\rho$ and $\phi$ are polar coordinates around this axis, and $2 \pi / \Omega$ is the pitch of the bundle.  Here, and below, I assume that $\Omega \rho$ is sufficiently small to neglect higher order corrections to the unit vector orientation.  As our central question focusses on the role of the twist-thermodynamics in selecting the size of bundles, we neglect the possibilities of anisotropic bundle cross-sections~\cite{hall2016, hall2017} as well as defects~\cite{grason2010, bruss2012, bruss2013} which can, in part, relax the cost of geometry frustration in the bundle.  Hence, the model considers bundles with a circular cross-section of radius $R$ and a length $L \ll R$, which is unlimited by equilibrium considerations.

Underlying the model described below are two basic assumptions.  First, filaments are sufficiently stiff, and interactions between them are sufficiently cohesive, that condensed bundles adopt quasi-parallel, splay-free packings with at a minimum, a nematic state of order.  Second, physical interactions between filamentous and columnar building blocks are treated at the mesoscopic scale of bundles by the generalized continuum elastic costs of gradients in the local order.  Hence, detailed properties of the inter-filament forces and intra-filament mechanics, as well as physical chemical parameters of the solution (e.g. temperature, ionic conditions), are incorporated in a coarse-grained sense into a limited set of continuum elastic constants.  The connection between microscopic descriptions of filaments and these mesoscale elastic constants is discussed briefly in Sec. \ref{sec: discuss}.

As shown schematically in Fig.~\ref{orderfig}, a bundle can be characterized by three types of elasticity associated with gradients of column {\it orientation}, {\it in-plane positional} order, and {\it longitudinal positional} order (e.g. for the case of intercalated discotic fibers).  The following subsections introduce the continuum energetics associated with each of these types of order.  As many of these elastic costs have been described elsewhere~\cite{grason2009, grason2012, hall2017}, hereonly a brief review of key results if given, citing previous work where possible, and relegating details of new analytical results in the appendices.

\subsection{Orientational elasticity}

To describe the free energy associated with orientational gradients in the bundle, consider the standard, second-order Frank elastic description~\cite{degennes_prost} of a chiral nematic materials with a director field, $\tv(\xv)$.  
\begin{multline}
F'_{nem} = \frac{1}{2} \int dV ~\Big\{ K_1 (\grad \cdot \tv)^2 + K_2 \big[ \tv \cdot (\grad \times \tv) + q_0 \big]^2 \\ + K_3 \big[ (\tv \cdot \grad) \tv \big]^2 + 2 K_{24}  \grad \cdot \big[ (\tv \cdot \grad) \tv  - \tv (\grad \cdot \tv) \big] \Big\} ,
\end{multline}
where the first three terms describe the respective splay, twist and bend elasticity, while the final term has been traditionally denoted as the ``saddle splay'' term~\cite{selinger2019}.  For a chiral material, $q_0 \neq 0$ parameterizes a preference for twist at linear order (i.e. with a preferred pitch of uniaxial cholesteric order, $2 \pi/q_0$).  It is straightforward to show that the double-twist texture of eq. (\ref{eq: tv}) gives
\begin{equation}
\grad \cdot \tv =0; \  \tv \cdot (\grad \times \tv) = 2 \Omega  ;\ (\tv \cdot \grad) \tv = - (\Omega^2 \rho ) \hat{\rho} 
\end{equation}
In the following, we will assume the volume integral can be split into an integral over cross-sectional area $dA$ (uniform up to rigid rotations) along length increments $dz$.  From this we can evaluate $F_{nem}$ (ignoring the effects of ends as  $L/R \to \infty$),
\begin{equation}
F'_{nem}/V = 2 K_2 q_0 \Omega+2(K_2 - K_{24})\Omega^2 +\frac{K_3 \Omega^4}{2} \langle \rho^2 \rangle
\end{equation}
where $\langle \rho^2 \rangle=A^{-1} \int dA~ \rho^2$ is the 2nd moment of the cross-sectional fiber area, equal to $\langle \rho^2 \rangle = R^2/2$ for cylindrical fibers of radius $R$. We note that the only quadratic terms in the double-twist derive from the twist and saddle splay.  Hence, as has been previously noted~\cite{koning2014, nayani2015, selinger2019} when the saddle splay constant is larger than the twist constant, $K_{24}>K_2$, an axisymmetric configuration becomes unstable to double-twist, and even the absence of chirality ($q_0 =0$), would be driven to spontaneous twist.  Several achiral liquid crystalline systems have been observed to undergo this spontaneous twist, and for the purposes of the present continuum model, I also consider the case of $K_{24}>K_2$ as a mesoscale mechanism for driving spontaneous twist in bundles, and reserve for the discussion the relationship between filament-scale interactions and these second order coefficients.   When the elasticity theory becomes unstable at second order in $\Omega$ it is necessary to include higher order gradient costs of twist which stabilize it.  For the present model, it is sufficient to generalize the nematic energy by $F_{nem} = F'_{nem} + (K'_2/2) \int dV ~ \big[ \tv \cdot (\grad \times \tv)\big]^4$, such that the cost for nematic gradients become,
\begin{equation}
\label{eq: fnem}
F_{nem}/V = 2 K_2 q_0 \Omega+2(K_2 - K_{24})\Omega^2 +\frac{K_3 \Omega^4}{4}R^2 +8 K_2' \Omega^4.
\end{equation}
The higher order twist term only becomes relevant in the limit of small bundles, in particular when $R \ll \sqrt{K_2'/K_3}$.  On dimensional grounds, it can be argued that the ratios is $K_2'/K_2$ (or $K_2'/K_3$) defines a length scale squared, and we argue below that this length scale defines the {\it microscopic cutoff} for the elasticity theory.  In other words, we expect that the length scale $\sqrt{K_2'/K_3}$ to be proportional to the microscopic dimension of filaments, e.g. their diameter $d$.  The implications of this microscopic dimension on bundle size-selection are described below.

\subsection{Positional elasticity}

\begin{figure*}
\centering
\includegraphics[width=0.8\linewidth]{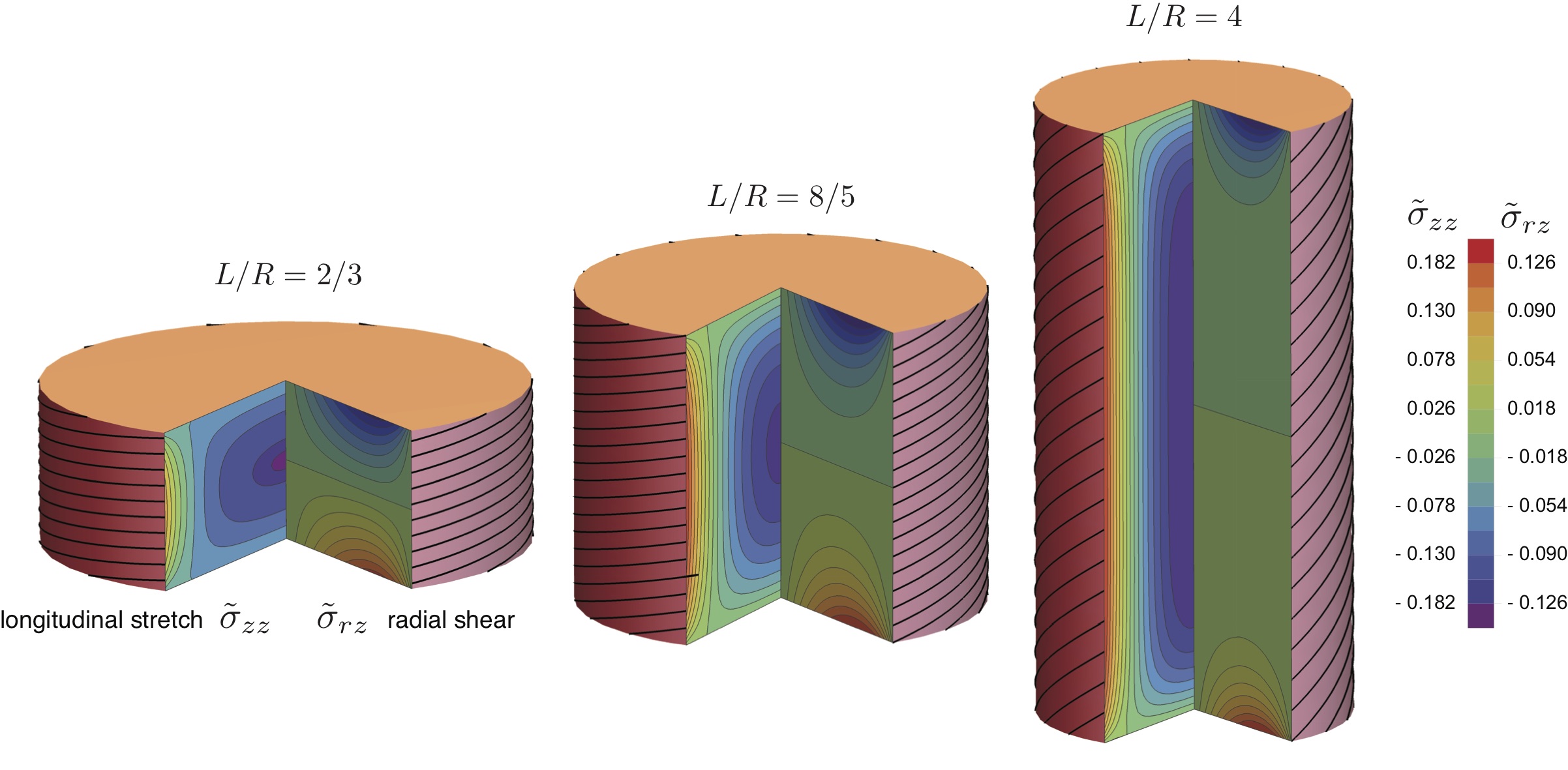}
\caption{Maps of stresses in twisted, 3D solid (crystalline) bundles plotted for three different bundle aspect ratios.  The cutaway shows the normalized intra-column stress ($\tilde{\sigma}_{zz} \equiv \sigma_{zz}/\big[ \lambda_\parallel (\Omega R)^2 \big] $) on the left, and the right shows the normalized radial shear ($\tilde{\sigma}_{rz} \equiv \sigma_{rz}/\big[ \sqrt{2 \mu_\parallel\lambda_\parallel} (\Omega R)^2 \big] $) .  Notably, away from the ends of long (and narrow) bundles, radial shear vanishes and intra-column stretching is nearly constant along the length.  Distributions are shown for the case $\lambda_\parallel = 2 \mu_\parallel$.}
\label{crystal}
\end{figure*}

Fiber assembly of 1D filaments of columns can give rise to different states of positional order.  Long-range ordering of the inter-filament spacing associate with a 2D lattice packing transverse to their backbones, but without long-range order along them, corresponds to {\it 2D columnar order}~\cite{chaikin_lubensky}.  Bundles that also maintain long-range axial correlations between neighbor columns/filaments (such as in an interdigitated lattice of 1D discoidal columns), correspond to {\it 3D crystalline order}.  Fiber twist generates deformations of both types of long-range positional order, and here I summarize the generalized positional elastic costs of cylindrical bundle twist, focussing first on 2D columnar order.

\subsubsection{Columnar order}

Elasticity of a columnar medium is described by a 2D strain tensor $u^\perp_{ij}$ for deformations perpendicular to the main filament axis, described by the elastic free energy~\cite{selinger_bruinsma1991},
\begin{equation}
F_{\perp} = \frac{1}{2} \int dV ~ \Big[ \lambda_\perp (u^\perp_{kk})^2 + 2 \mu_\perp ~ u^\perp_{ij}u^\perp_{ij} \Big],
\end{equation}
where the (non-linear) 2D elastic strain follows,
\begin{equation}
u_{ij}^\perp \simeq \frac{1}{2} \big( \partial_i u^\perp_j +  \partial_j u^\perp_i - t_i t_j \big)
\end{equation}
where ${\bf u}_\perp$ is the in-plane displacement (2D vector), related to the tangent field via $\tv \simeq \hat{z} + \partial_z {\bf u}_\perp$.    The Lam\'e coefficients $\lambda_\perp$ and $ \mu_\perp$ describe, at a coarse-grained level, the disruptions of the ideal 2D inter-column lattice, which we assume to be hexagonal for simplicity.  The non-linear contribution to strain from the in-plane projections of the filament tilt generates unavoidable inter-column stress, $\sigma^\perp_{ij}$, as has been shown~\cite{grason2012} to derive from the compatibility condition,
\begin{equation}
\grad^2_\perp \sigma^\perp_{ii} = - 3 Y_\perp \Omega^2
\end{equation}
for non-zero twist, where $Y_\perp= 4 \mu_\perp ( \lambda_\perp + \mu_\perp)/ ( \lambda_\perp + 2\mu_\perp)$ is the 2D Young's modulus of the columnar array.  Solving for equilibrium stress for a bundle of circular cross section of radius $R$ the in-plane elastic free energy has been derived~\cite{grason2012},
\begin{equation}
\label{eq: fper}
F_{\perp}/V= \frac{3 Y_\perp (\Omega R)^4}{128} .
\end{equation} 
This term represents the elastic cost of geometric incompatibility of a crystalline packing with metric constraints imposed in non-parallel and twisted bundles~\cite{grason2015}.  Note that as the 2D columnar order melts, the resistance to shear of the inter-filament lattice vanishes.  Hence, the limit $Y_\perp \to 4 \mu_\perp \to 0$ corresponds to the transition from columnar to polymer nematic order in the bundle.

\subsubsection{Crystalline order}

Now I consider the additional elastic costs of twist in  bundles whose columns maintain registry of longitudinal stacking (i.e., 3D crystalline).  The additional deformations associated with longitudinal inter-column shears and intra-column stretch are described by the out-of-plane elastic energy~\cite{landau1986},
\begin{equation}
F_{\parallel} = \frac{1}{2} \int dV ~\Big[ \lambda_\parallel u_{zz}^2 + 2 \mu_\parallel (u_{xz}^2 + u_{yz}^2 ) \Big] ,
\end{equation}
where $u_{ij}$ is the 3D solid elastic strain tensor (with components in $x, y$ and $z$ directions),
\begin{equation}
u_{ij} = \frac{1}{2} \big( \partial_i u_j + \partial_i u_j + \partial_i  {\bf u} \cdot  \partial_j  {\bf u} \big)
\end{equation}
where ${\bf u}= {\bf u}_\perp + u_z \hat{z}$ is the 3D displacement of column positions relative to a parallel, hexagonal reference state~\footnote{Here, I drop a term proportional $u_{zz} (u_{xx} + u_{yy})$ since this will lead to a small renormalization of the energetic term proportional to column stretching $\lambda_{\parallel}$}.  Here $\lambda_\parallel$ and $\mu_\parallel$ parameterize the respective stretching elasticity of columns and inter-column shear coupling. The equilibrium equation for the longitudinal displacement $u_{z}$ is 
\begin{equation}
\partial_i \sigma_{iz} =0; \ \sigma_{zz} =  \lambda_\parallel \big(\partial_z u_z +\Omega^2 r^2/2 \big) ; \ \sigma_{i z} =  \mu_\parallel \big( \partial_i u_z + t_i) .
\end{equation}
The full solution to these equations is given in Appendix~\ref{crystalline}, but here I summarize the essential results, first focusing on the case of a finite fiber length $L$.  The solid response of bundles to twist derives from three deformations:  i) intra-filament stretch; ii) {\it radial} inter-filament shears; and iii) {\it azimuthal} inter-filament shears.  While azimuthal shears are unavoidable for all twisted bundles (i.e. $\sigma_{\phi z} \neq 0$), sufficiently {\it long} bundles avoid the cost of radial shears at the expense of filament stretch.  This deformation crosses over to a shear-dominated state for sufficiently short bundles, with a cross-over determined by the ratio $\sqrt{ \mu_\parallel/\lambda_\parallel} (L/R)$.

In the limit of narrow (long) fibers $R \ll ( \mu_\parallel/\lambda_\parallel)^{1/2}  L$, the solution (except for a small boundary layer at the fiber ends) becomes $u_z=0$, leading to a lateral stretching between shear-coupled filaments, $u_{zz} \to \Omega^2(r^2-R^2/2)/2$, which is zero net tension when averaged over the bundle cross section.  In the opposite limit of short (wide) fibers $R \gg  ( \mu_\parallel/\lambda_\parallel)^{1/2}  L$, the cost of filament stretching becomes prohibitive and the equilibrium tends towards inextensible $u_{zz} \to 0$ and $u_z \to - z \Omega^2 r^2/2$, leading to radial shears $u_{rz} \approx  \Omega^2 z  r$ that grow with length.  The crossover from shear to stretch dominated stress with increasing aspect ratio is illustrated for stress profiles in Fig.~\ref{crystal}.  For both cases, twist generates azimuthal shears $u_{\phi z} = \Omega r/2$.  Together these lead to the strain energy dependence on twist,
\begin{equation}
\label{eq: fpar}
F_{\parallel}/V=\left\{\begin{array}{ll}\frac{ \mu_\parallel }{8} ( \Omega R)^2+ \frac{\lambda_\parallel }{24}   (\Omega R)^4, & R \ll  ( \mu_\parallel/\lambda_\parallel)^{1/2}   L \\ \\ \frac{ \mu_\parallel }{8} ( \Omega R)^2+  \frac{ \mu_\parallel }{96} \Omega^4 R^2 L^2, & R \gg  ( \mu_\parallel/\lambda_\parallel)^{1/2}   L \end{array} \right.
\end{equation}
The Appendix -- eqs.(\ref{eq: stretch}) and  (\ref{eq: shear}) -- give the exact result that crosses over continuously from the shear dominated to stretch dominated regime with decreasing values, $\sqrt{ \mu_\parallel/\lambda_\parallel} (L/R)$.  Notably $F_{\parallel} \to 0 $ as shear coupling between columns vanishes $\mu_\parallel \to 0$, leaving only the 2D elastic (columnar) terms. 

Below, I consider only the case of narrow bundles, $R \ll ( \mu_\parallel/\lambda_\parallel)^{1/2}  L$, relevant to fibers that assemble end-to-end without constraint on lengths (i.e. $L \to0$ while $R$ may remain finite).  This twist dependent costs of intra-filament stretch in 3D crystalline fibers was first proposed to limit their diameter in works by Makowski~\cite{makowski1986} and Weisel~\cite{weisel1987}, followed by more detailed analytical models in ref. ~\cite{turner2003} .  Here, it should be noted that these previous studies neglected the unavoidable costs of azimuthal shears, which unlike the radial shear cannot be relaxed by longitudinal displacement.  Eq. (\ref{eq: fpar}) shows that azimuthal shear in solid bundles generates second order costs for twist, growing as $(\Omega R)^2$, well known in mechanics for the twist elasticity of solid beams~\cite{landau1986}, and more important, at lower order than the quartic intra-filament stretching energy.

\begin{figure*}
\centering
\includegraphics[width=1\linewidth]{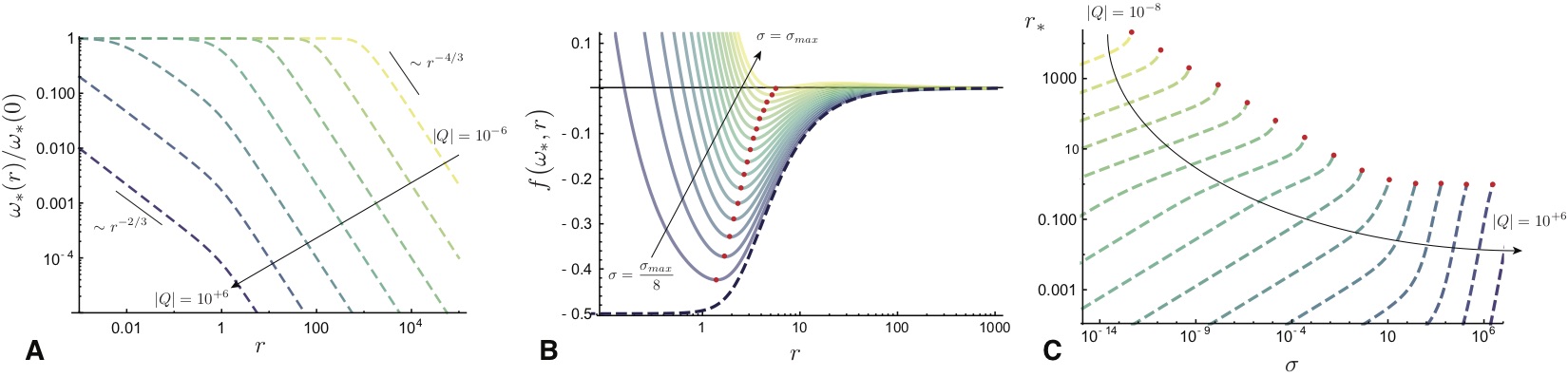}
\caption{Thermodynamic behavior of {\it chirality-driven, twist-stable 2D columnar bundles} (i.e. $Q\neq 0; \chi^{-1}>0; \mu_\parallel = 0$).  (A) shows the equilibrium twist $\omega_*$ as a function bundle radius $r$, normalized by the value of $r\to 0$ bundles.  Behaviors for a series of chiralities $Q$ different by factors of $10^2$. (B) shows plots of the energy density for equilibrated twist (i.e. $\omega = \omega_*$) versus bundle radius $r$ for fixed chirality $|Q| = 0.1$, plotted for a series of increasing surface energy $\sigma$ up to a maximal value of $\sigma_{max}$, at which point finite bundles are in equilibrium with bulk untwisted assembly (i.e. $r \to\infty$ and $\omega \to 0$).  The accumulating elastic energy of bend and columnar strain is shown as a dashed blue curve, and equilibrium bundle radii $r_*$ are marked by red dots.    In (C), plots of equilibrium radii $r_*$ versus surface energy, $\sigma$, for an increasing series of chirality values.  Curves terminate at the $\sigma_{max}$, with the maximal self-limiting size shown as red dots.  Plots in (A-C) are shown for $r_0 = 0$ and $\chi^{-1}=1$.   }
\label{chiral_col}
\end{figure*}

\subsection{Total free energy and reduced variables}

The total free energy is constructed from the terms described above along with a surface energy cost of $\Sigma$ per unit area of the cylindrical sides of the bundles,
\begin{equation}
F_{tot}= F_{nem}+F_{\perp} + F_{\parallel} + 2 \pi R L \Sigma.
\end{equation}
In what follows, the thermodynamics of finite-size, twisted bundles is more conveniently analyzed by rescaling energy densities in terms of the effective positional modulus,
\begin{equation}
Y\equiv \frac{3}{32} Y_\perp + \frac{1}{6} \lambda_\parallel ,
\end{equation}
and length scales in terms the ratio of bending modulus to positional modulus 
\begin{equation}
\Lambda_B = \sqrt{ K_3/Y} ,
\end{equation}
a quantity related to the {\it bend penetration length} of columnar systems.  Note, in terms of the this length scale, the transition from 2D columnar to purely nematic bundles is characterized by $\Lambda_B \to \infty$, since $Y$ vanishes as lattice order melts.

In terms of the reduced bundle radius $r = R/\Lambda_B$ and twist $\omega = \Omega \Lambda_B$, the reduced free energy density $f$ may be written in the following general form,
\begin{equation}
\label{eq: f}
f(\omega, r) \equiv \frac{F_{tot}}{Y V} = Q \omega + \frac{\chi^{-1}(r) }{2} \omega^2 + \frac{\beta(r)}{4}\omega^4 + \frac{\sigma}{r} ,
\end{equation}
where $\sigma \equiv 2 \pi \Sigma / \sqrt{Y K_3}$ is the reduced surface energy.   In terms of the generalized elastic theory, the {\it reduced chirality} is simply $Q = 2 K_2 q_0/\sqrt{Y K_3}$, which is proportional to the preferred inverse cholesteric pitch.  The {\it inverse twist susceptibility} has the general form,
\begin{equation}
\label{eq: chiinv}
\chi^{-1} = \chi^{-1}_0 + \chi_2^{-1} r^2 ,
\end{equation}
where $\chi^{-1}_0 = 2 (K_2- K_{24})/K_3$ derives from the twist elasticity of orientational (nematic) order, while the size-dependent contribution derives from the shear-elastic cost in solid bundles, $\chi_2^{-1} = \mu_{\parallel}/(8Y)$. Notably, this shows that twist instability  ($\chi^{-1}<0$), which arises for large saddle-splay constants, is only possible in 3D solid bundles of sufficiently narrow radius due to size-dependent costs of inter-filament shears.  The coefficient of the quartic twist term grows with radius,
\begin{equation}
\beta(r) = r_0^2 + r^2 +r^4
\end{equation} 
due to the respective $r^2$ and $r^4$ costs of bending and positional elasticity.  The constant term $r_0 \equiv 4 \sqrt{ 2 K'_2/K_3}/\Lambda_B$ derives from the higher-order twist in the nematic energy, eq. (\ref{eq: fnem}).  It can be argued that the ratio of bend to positional elasticity in a filament bundle gives bend penetration length that is at least as large as the microscopic inter-filament dimension, $d$~\footnote{This follows from a simple estimate of the intra-filament elasticity, in terms of a solid elastic modulus $E$.  The bending modulus of a single filament follows beam mechanics~\cite{landau1986}, $B \propto E d^4$, such that $K_3 \approx E d^2$.  The inter-filament modulus $Y_\perp$ is determined by the softer of either inter-filament cohesion, or the intra-filament deformabilty itself.  Hence, $Y_\perp \lesssim E$ (note that $\lambda_\parallel \approx E$).  Hence, this gives $K_3/Y \gtrsim d^2$.} , that is, $\Lambda_B \gtrsim d$.  Again, taking the estimate $4\sqrt{ 2 K'_2/K_3} \approx d$ then gives us that $r_0 \lesssim 1$, a parameter estimate used in the analysis below.

\section{Thermodynamics of twist- and size-selection}

\label{sec: thermo}

Based on the model introduced in the previous section and summarized in the scaled free energy in eq. (\ref{eq: f}), I now describe the thermodynamics of self-assembled, twisted fibers.  Here, consider the case where the total concentration of subunits is sufficiently large that all but a negligible concentration exists in a self-assembled aggregate.  In this regime, thermodynamics can be modeled by considering all filaments assembled into bundles of equal size $r_*$ and twist $\omega_*$ whose values correspond to the minimum of the free energy density $f(\omega,r)$, that is, neglecting the effect of size dispersity on the free energy of the distribution (see e.g. \cite{groenewold2001}).

To address the central questions about the role of {\it twist-stability} vs. {\it twist-instability} on bundle formation, I describe three inter-related behaviors for columnar (2D solid) and crystalline (3D solid) bundles.  First, I describe the dependence of equilibrium twist $\omega_*(r)$ on bundle size $r$ and its relation to the accumulation of elastic twist energy with radius.  Second, I show accumulating elastic energy in self-twisting bundles leads to self-limited equilibrium radii $r_*$ for sufficiently low surface energy $\sigma$.  In general, $r_*$ increases with $\sigma$ up to a maximum self-limiting size, $r_{max}$ and surface energy $\sigma_{max}$, beyond which surface energy drives equilibrium states to untwist and reach infinite size.  Last, I describe how this maximal size and surface energy of stable finite-width bundles varies with chirality and twist susceptibility.  In what follows, these results are illustrated in Figs.~\ref{chiral_col} -~\ref{Figcryst_rmax},  with details on the numerical analysis of the equilibria of $f(\omega, r)$ provided in Appendix~\ref{generalsol}.

\subsection{Columnar bundles}

Considering first the case of columnar bundles, which corresponds to a twist susceptibility that is independent of bundle size, $\chi = \chi_0$.  For purposes of illustration, I highlight the comparison between {\it chirality driven, twist-stable} bundles ($Q\neq 0; \chi^{-1}>0$) to {\it spontaneously twisting, achiral} bundles ($Q=0;  \chi^{-1}<0$), and then summarize the general dependence of self-limiting assembly on $Q$ and $\chi$.

\subsubsection{Chirality-driven twist}
Figure \ref{chiral_col}A plots the equilibrium twist $\omega_*$ as a function of bundle radius $r$ for several examples of chirality-driven, twist-stable bundles.  For twist-stable bundles, the higher order twist contribution to the Frank elastic energy plays a relatively minor role in the qualitative behavior, and hence I set it to $r_0 =0$ for these examples.  These curves all show a maximal twist in the limit of narrow bundles proportional to reduced chirality, $\omega_*(r \to 0) = - \chi Q$, as there is no mechanical obstruction to achieving the double-twist state preferred by nematic twist elasticity.  As bundle sizes increase, the equilibrium begins to unwind from this preferred value due to the mechanical costs of intra-filament bending and inter-filament lattice distortion.  This unwinding of helical pitch can be characterized by a size scale $r_{un}$, the {\it unwinding size}, at which the elastic cost of either bending or lattice distortion (i.e. $(Q \chi)^4 r^2/4$ or $ (Q \chi)^4  r^4/4$, respectively) equals and begins to exceed the favorable energy of chiral twist at the preferred pitch, $-Q^2 \chi/2$.  Based on this criterion the unwinding size is roughly $r_{un} \approx {\rm min} \big[ (Q^2 \chi^3)^{-1/2}, (Q^2 \chi^3)^{-1/4} \big]$, and allows us to distinguish between two regimes of chirality:  {\it weak chiralty} where $Q^2\ll \chi^{-3}$ and $r_{un} \gg 1$; and {\it strong chirality} where $Q^2\gg \chi^{-3}$ and $r_{un} \ll 1$.  For larger sizes, $r \gg r_{un}$, bundles unwind toward zero twist, and the rate of unwinding with size can be estimated from the balance of chirality induced torque $Q$ and the torque induced by mechanical costs of twist, $\omega_*^3 (r^2 + r^4)$, which yields a power-law unwinding that crosses over from bending dominated unwinding $\omega_* \sim r^{-2/3}$ for $r_{un} \ll r \ll 1$ to a more rapid columnar strain-dominated unwinding $\omega_* \sim r^{-4/3}$ at larger sizes $r\gg 1$.  

The free energy density of twist equilibrated bundles, that is, taking the value $\omega = \omega_*(r)$ in the $f(\omega, r)$, is plotted in Figure \ref{chiral_col}B for a fixed chirality and an increasing series of surface energies $\sigma$.  For $\sigma=0$, the dashed line shows the monotonoically increasing accumulation of elastic energy with size.  Narrow bundles achieve the optimal twist without the expense of bending or columnar strain costs.  At intermediate sizes, the elastic energy exhibits a power law growth (either as $\sim r^2$ or $\sim r^4$) which then crosses over to an asymptotically unwinding state with $f(r \gg r_{un}) \to 0$.  While the elastic energy generically favors narrow bundles, the (per volume) cost of surface energy, $\sigma/r$, drives bundles to larger sizes.  The balance between elastic energy and surface energy results in an equilibrium at finite size $r_*$ for sufficiently small $\sigma$  As surface energy increases, the size and energy of this minimum grow, until it reaches a point where $f(\omega_*, r_*) = 0$ at $\sigma_{max}$ and finite size bundles are in equilibrium with bulk, untwisted assembly.  A narrow range of metastable finite bundles persists above this surface energy, but the equilibrium state for $\sigma > \sigma_{max}$ is bulk, untwisted assembly (i.e. $\omega \to 0$ and $r \to \infty$).

\begin{figure*}
\centering
\includegraphics[width=1\linewidth]{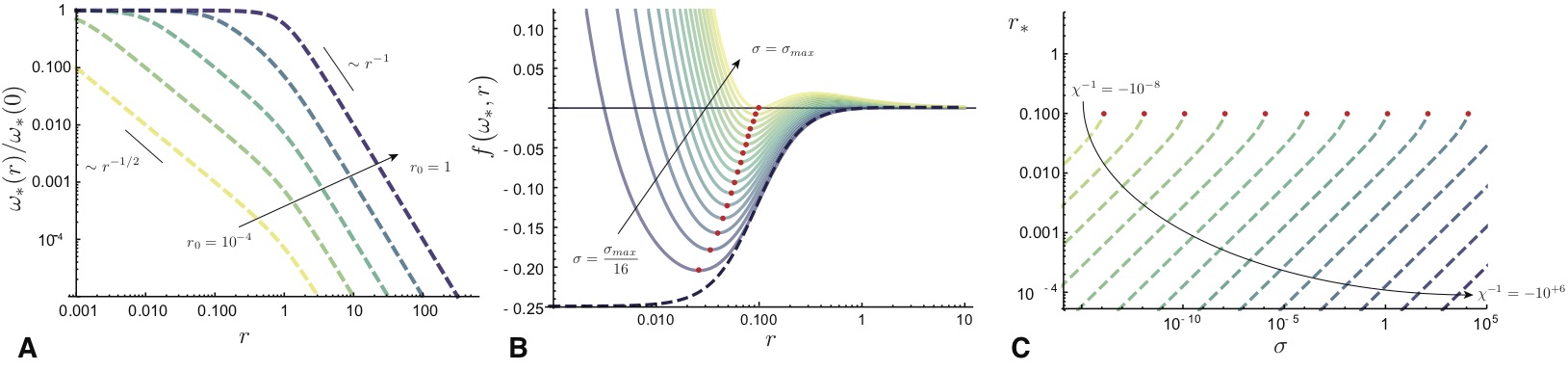}
\caption{Thermodynamic behavior of {\it spontaneously-twisting, achiral 2D columnar bundles} (i.e. $Q\neq 0; \chi^{-1}<0; \mu_\parallel = 0$).  (A) shows the equilibrium twist $\omega_*$ as a function bundle radius $r$, normalized by the value of $r\to 0$ bundles.  Behaviors are plotted for fixed $\chi^{-1} = -1$ and a range of ``cutoff'' size scales $r_0= 10^{-4},10^{-3},10^{-2},10^{-1}$ and 1.  (B) shows of the energy density for equilibrated twist (i.e. $\omega = \omega_*$) versus bundle radius $r$ for fixed $r_0 = 0.1$ and $\chi^{-1} = -1$, plotted for series of increasing surface energy $\sigma$ up to a maximal value of $\sigma_{max}$, at which point finite bundles are in equilibrium with bundle untwisted assembly.  The accumulating elastic energy of bend and columnar strain is shown as a dashed blue curve, and equilibrium bundle radii $r_*$ are marked by red dots.  In (C), plots of equilibrium radii $r_*$ versus surface energy for increasingly negative values of $\chi^{-1}$.  Curves terminate at $\sigma_{max}$, with the maximal self-limiting size shown as red dots.}
\label{achiral_col}
\end{figure*}

The equations of state, relating equilibrium radius $r_*$ to surface energy $\sigma$ are plotted for a sequence of chirality values in Figure \ref{chiral_col}C, with the curves terminating a the maximal size and surface energy for self-limitation.  The size dependence exhibits two regimes of assembly: weak-chirality when $|Q| \ll \chi^{-3/2}$; and strong-chirality when $|Q| \gg \chi^{-3/2}$.  For strong-chirality the bundles begin to unwind (due to bending) for sizes well below the mesoscopic length scale $\Lambda_B$.  In the bending dominate regime, the residual free energy from chiral twist $\approx -\omega_* Q \sim -r^{-2/3}$ vanishes more slowly than the surface energy cost $\sigma/r$ for large bundle radii, implying the existence of a stable equilibrium size with the power law dependence $r_* \sim \sigma^3$.  However, when bundles grow larger than $\Lambda_B$ (i.e. when $r\gg 1$) the cost of columnar strain drives a much more rapid untwisting such that the chiral free energy $\approx -\omega_* Q \sim -r^{-4/3}$ cannot restrain the stronger $\sim r^{-1}$ dependence of surface energy for $r \to \infty$, indicating the disappearance of the stable minimum at finite size for $r_* \gtrsim 1$ for strong-chirality.  This is consistent the $r_{max} \to 1$ behavior for large $Q$ shown in Figure \ref{chiral_col}C.

For weak-chirality, $r_{un} \gg 1$, so that the bundles accumulate stress over a large size range, well into the lattice-strain dominant regime, before untwisting.  Balancing the surface energy cost $\sigma/r$ of the boundary with the dominant elastic cost results in two power law regimes of bundle size below the untwisting size:  $r_* \approx \sigma^{1/3} (Q \chi)^{4/3}$ for bending dominated sizes ($r_* \ll 1$); and $r_* \approx \sigma^{1/5} (Q \chi)^{4/5}$ for lattice-strain dominated sizes ($r_* \gg 1$).    Note that, for the same arguments as above, because the columnar strain energy forces the bundles to unwind too rapidly to be at finite size in equilibrium surface energy, there can be no stable equilibrium sizes larger than $r_{un}$ in this weak-chirality regime.  Hence, the power law growth of $r_*$ with $\sigma$  persists until terminating at a maximum size, $r_{max} \approx r_{un} \sim |Q|^{-1/2} \chi^{-3/4}$, consistent with the low $|Q|$ scaling of $r_{max}$ in Figure \ref{chiral_col}C.

To summarize, for chirality-driven bundles, the maximal size of self limiting bundles follows
\begin{equation}
r_{max} \approx \left \{ \begin{array}{ll} |Q|^{-1/2} \chi^{-3/4}, & {\rm for} \ |Q| \ll \chi^{-3/2}; \chi^{-1}>0  \\ 1, &  {\rm for} \ |Q| \gg \chi^{-3/2}; \chi^{-1}>0 \end{array} \right.
\end{equation}
Notably, for chirality-driven, twist-stable bundles, finite-size assembly always extends up to at least the mesoscopic size $\Lambda_B$ (i.e. $r_* \geq 1$), and as chirality decreases, the maximal stable finite bundle size diverges, $r_{max} \sim |Q|^{-1/2} \gg 1$, growing arbitrarily larger than  $\Lambda_B$ as $Q \to 0$.  

\begin{figure*}
\centering
\includegraphics[width=0.9\linewidth]{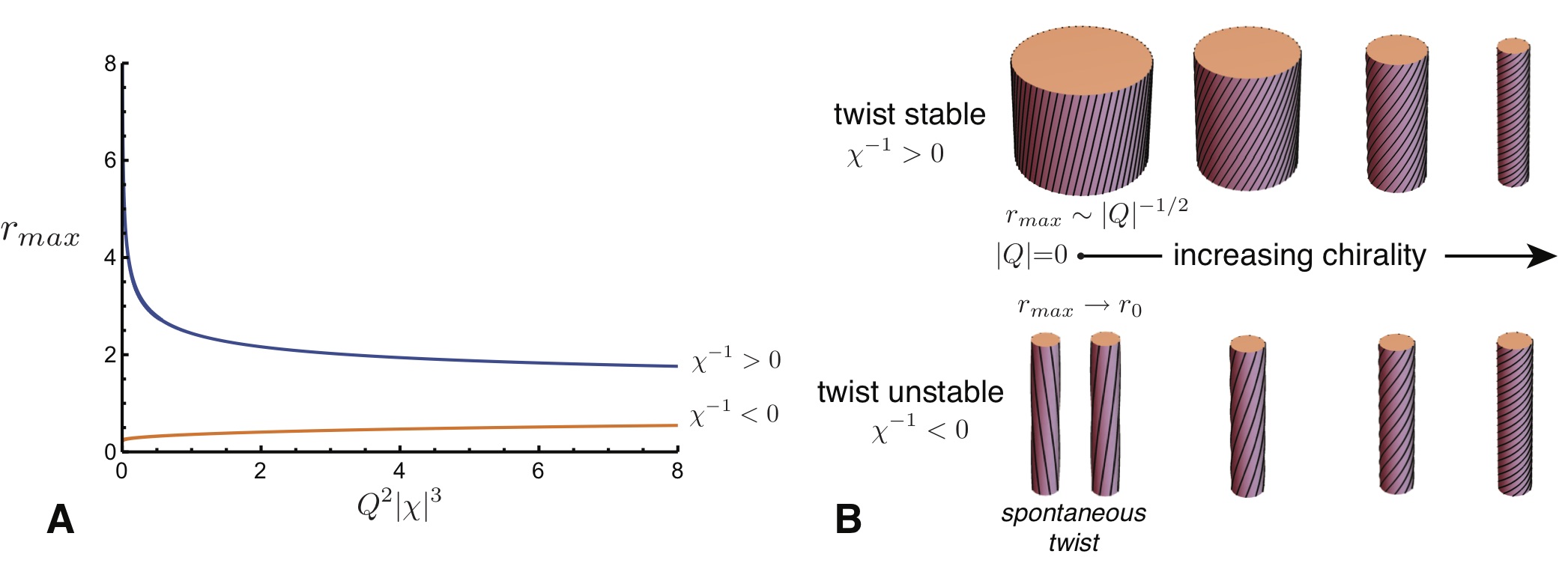}
\caption{In (A), plot of maximal self-limiting bundle radius ($r_{max}$) plotted for arbitrary chirality ($Q$) and inverse twist-susceptibility ($\chi^{-1}$), showing it to be a function of a combination of the two parameters, $Q^2 \chi^3$.   The {\it twist-stable} case ($\chi^{-1}>0$) and the {\it twist-unstable} case ($\chi^{-1}<0$) are plotted in blue and orange curves, respectively.  The distinct dependencies on chirality of the maximal size self-limited bundles (for fixed $\chi^{-1}$) are illustrated schematically in (B), with {\it twist stable} bundles growing arbitrarily large in the small chirality limit, while {\it twist-unstable} bundles remain smaller than the length scale $\Lambda_B$ (i.e. $r_{max} \leq 1$) over the entire chirality range.  At the achiral point $Q=0$, highlighted for {\it twist-unstable} bundles, the equilibrium handedness of finite bundles is randomly selected (i.e. by spontaneous achiral symmetry breaking).}
\label{rmax_col}
\end{figure*}

\subsubsection{Spontaneous twist}

Figure \ref{achiral_col} shows plots of the thermodynamic behavior of spontaneously twisting, achiral bundles, paralleling the presentation of Fig. ~\ref{chiral_col}.  For this case of unstable twist, $\chi^{-1} < 0$, it is necessary to include the effect of the higher-order Frank twist term, as parameterized by a non-zero ``cutoff size scale" $r_0 \neq 0$, which is anticipated to be less unity as argued above.   Figure \ref{achiral_col}A plots equilibrium twist a function of bundle radius, in this case for several values of $r_0 \leq1$.  In this achiral case, the maximal twist achieved in narrow bundle limit is $\omega_*( r\to 0) =\pm |\chi|^{-1/2}/ r_0$.  Like the chirality-driven case, spontaneously twisting bundles also exhibit a size-dependent unwinding due to the accumulated costs of bending.  Applying similar reasoning as above, the unwinding size can be estimated as $r_{un} \approx r_0$~\footnote{Note that the assumption that $r_0 \leq 1$ implies that unwinding is always bending-dominated.}. Beyond this size, bending and columnar strain induce the unwinding of bundles towards $\omega_* \to0$ as $r \to 0$.  While qualitatively similar to the unwinding of chirality-drive bundles, spontaneously twisted bundles unwind much more rapidly with increases size:  $\omega_* \sim r^{-1}$ in the bending dominated regime ($r_{un} \ll r \ll 1$); and $\omega_* \sim r^{-2}$ in the columnar strain dominated case ($r \gg 1$).  The origin of these much stronger power laws can be traced to the torque induced by the spontaneous twist, which vanishes for as $\omega \to 0$ as $\sim - |\chi^{-1}|\omega$, yeilding a much weaker resistance to the mechanical costs that drive unwinding.  I return to the implications of the more rapid unwinding of spontaneously twisting bundles below.
 
Figure \ref{achiral_col}B plots the free energy of twist-equilibrated bundles, $f(\omega_*,r)$ , as function of bundle size for $\chi^{-1} = -1$, $r_0 =0.1$ and a series of surface energies up to maximum surface energy $\sigma_{max}$.  Qualitatively, the features of the size-dependent energy for achiral bundles parallel what was shown for chiral bundles in Fig. \ref{chiral_col}B:  accumulating elastic costs of frustration balances the surface energy drive towards bulk assembly, but with finite sizes of minimal achiral bundles are notably smaller, i.e. $r_*< 1$.

The dependence of equilibrium size on surface energy is plotted in Figure \ref{achiral_col}C for $r_0=0.1$ and for an increasing series of $\chi^{-1}=-1$, corresponding to increasing amounts of spontaneous twist (i.e. $\omega_*( r\to 0) \propto  \sqrt{-\chi^{-1}}$).  The equilibrium bundle radius grows with as $r_* \sim \sigma^{1/3}$ due to a balance between surface energy and bending of spontaneously twisted filaments.  All of these curves terminate at  maximal size $r_{max} \approx r_0$, which derives from the fact twist unwinds at size scale $r_{un} \approx r_0$.  In the unwound, bending-dominated regime, the residual energy of spontaneous twist falls of as $\approx \chi^{-1} \omega_*^2 \sim - r^{-2}$, too rapidly in comparison to surface energy to maintain a self-limited equilibrium in this large (unwinding) size regime.  

To summarize, in sharp distinction which chirality-driven, twist-stable bundles, the range of self-limited sizes of spontaneously twisted achiral bundles is limited to a much smaller size range,
\begin{equation}
r_{max} \approx r_0, {\rm for \ all \ } \chi^{-1}<0;  Q=0 .
\end{equation}
As $r_0$ is expected to be comparable to the microscopic dimensions of the filament or column diameter, this prediction shows that while spontaneously, twisting bundles can (at sufficiently low surface energy) realize self-limiting assembly, their sizes are restricted to a microscopic range of a few filaments in width.

\subsubsection{Phase diagram and maximal self-limiting size}
The previous sections have presented, in detail, the specific cases of chirality-driven, twist-stable bundles ($Q \neq 0; \chi^{-1}>0$) and spontaneously-twisted, achiral bundles ($Q=0; \chi^{-1} <0$).  In this section, I describe the generic thermodynamic behavior for generic ranges of chirality and twist susceptibility, in terms of the maximal size and surface energy for self-limited columnar bundles.  As derived in the Appendix, these conditions can be captured in the following parametric relationship between the $r_{max}$, $\sigma_{max}$, $Q$ and $\chi^{-1}$ (for columnar order):
\begin{equation}
\label{eq: sigmax}
\sigma_{max} \chi^2 = \frac{ 2 r_{max} (2 r_{max}^4+r_{max}^2)}{\big(r_{max}^4-r_{max}^2-3r_0^2\big)^2 } ,
\end{equation}
and
\begin{equation}
\label{eq: chirality}
Q^2 \chi^3 = \frac{ 2  \big(3 r_{max}^4+r_{max}^2-r_0^2\big)^2}{\big(r_{max}^4-r_{max}^2-3r_0^2\big)^3 } .
\end{equation}
Figure ~\ref{rmax_col}A shows the variation of the maximal self-limiting bundle radius as function of $Q^2 \chi^3$ for both the {\it twist-stable} and {\it twist-unstable} branch, with the predictions illustrated graphically in Figure ~\ref{rmax_col}B.  Notably, $r_{max}$ is a decreasing function of chirality (for fixed $\chi^{-1} >0$) for twist-stable bundles: The self-limiting size diverges as $r_{max} \sim |Q|^{-1/2}$ in the limit of vanishing chirality, while it asymptotically approaches $r_{max} \to 1$ in the limit of large chirality (assuming that $r_0 \lesssim 1$).  Heuristically, this can be understood from the fact that increasing chirality {\it decreases} the unwinding size, due to the increase of elastic cost with larger twist.  In contrast, the maximal radius of {\it twist-unstable}  bundles increases, but only very weakly, with chirality (for fixed $\chi^{-1} <0$):  In the achiral limit $r^2_{max} \to (\sqrt{1+12 r_0^2} -1)/6 \simeq r_0^2$, while for large chirality $r_{max} \to 1$ from below.  The relative insensitivity of $r_{max}$ to chirality for twist-unstable bundles can be understood from the achiral case illustrated in Fig. ~\ref{achiral_col}, where the degree of twist, and the unwinding size, is set by $r_0$ and not $Q$.  Hence, the self-limiting thickness of {\it twist-stable} bundles varies over a large mesoscopic range with chirality, while by comparison, the self-limiting thickness of {\it twist-unstable} bundles is both much smaller and varies relatively little with even large changes in chirality.  

\begin{figure}
\centering
\includegraphics[width=0.9\linewidth]{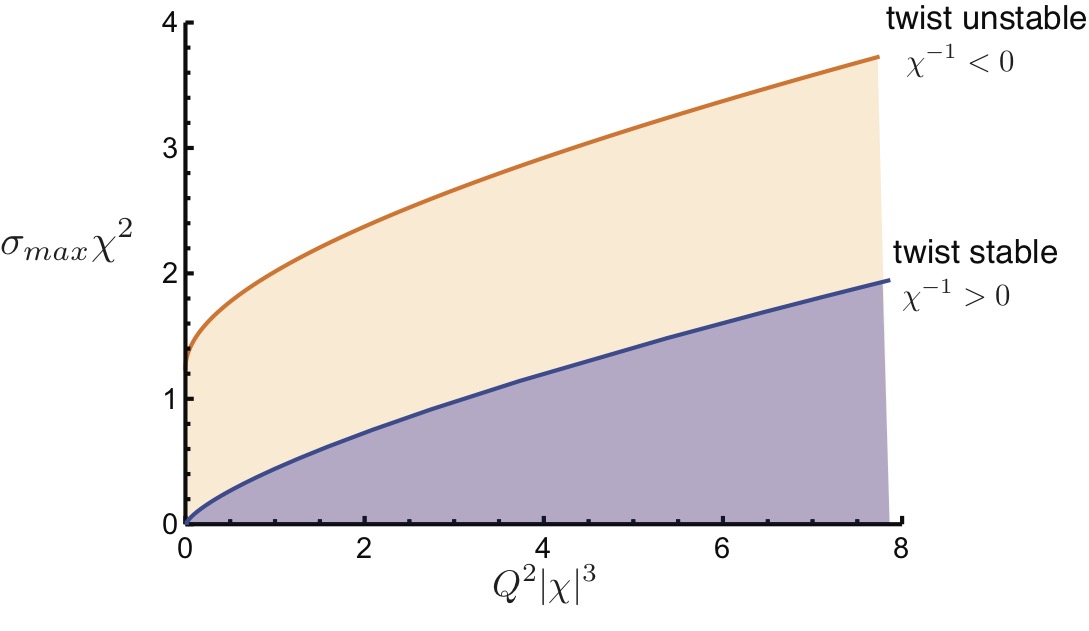}
\caption{Phase diagram of self-limiting assembly in {\it twist-stable} and {\it twist-unstable} columnar bundles.  For a given value of $Q^2 |\chi|^3$, finite-radius bundles are stable for $\sigma \leq \sigma_{max}$ (indicated as the shaded blue and orange regions) and for $\sigma> \sigma_{max}$ (white region) the equilibrium phase is bulk, untwisted assembly.}
\label{sig_col}
\end{figure}

Figure ~\ref{sig_col} shows the phase diagrams of self-limiting bundle assembly for both twist-stable and twist-unstable bundles: $\sigma_{max}$ as a function of $Q^2 |\chi|^3$.  For $\sigma < \sigma_{max}$ the equilibrium phase is characterized by finite-radius and finite-twist bundles, whereas $\sigma > \sigma_{max}$ the equilibrium phase is bulk, untwisted assembly ($r_* \to \infty, \omega_* \to 0$).  Increasing chirality at fixed $\chi^{-1}$ increases the stability range of finite bundles for both twist-stable and twist-unstable bundles, consistent with the $\sigma_{max} \sim Q^{2/3}$ scaling for large $Q$ in eqs. (\ref{eq: sigmax}) and (\ref{eq: chirality}).  This derives simply from the fact that higher chirality generically increases the free energy difference between locally twisted and untwisted assembly.  However, there is a notable difference in the vanishing chirality limit, deriving from the obvious distinction that twist-stable bundles require chirality to twist:  $\sigma_{max} =0$ as in the achiral limit of twist-stable bundles; while $\sigma_{max}$ remains finites as $Q \to0$ for twist-unstable bundles.  More generally, beyond the achiral limit, $\sigma_{max}$ is always greater for twist-unstable bundles than for twist-stable bundles.  Thus, while they exhibit a far smaller range of possible self-limiting sizes, self-limited bundles formed in {\it twist-unstable} assemblies exhibit a greatly enhanced range of thermodynamic stability relative to {\t twist-stable} bundles of equal chirality.

\subsection{Crystalline bundles}

In this section, I briefly overview the thermodynamics of 3D crystalline bundles.  These are distinguished from the 2D columnar case through the presence of a non-zero elastic cost for azimuthal shears (i.e. inter-column sliding) in twisted bundles, as characterized by $\mu_{\parallel} >0$ in eq. (\ref{eq: fpar}).  In terms of the dimensionless inverse twist susceptibility, $\chi^{-1} = \chi_0^{-1} +r^2  \chi_2^{-1}$, this corresponds to $\chi_2^{-1} \neq 0$, and an increasing twist stiffness with lateral size.   From the definition in eq. (\ref{eq: chiinv}), it can be shown that $\chi_2^{-1}$, which we denote as the {\it reduced solid modulus}, is strictly less than $\mu_{\parallel}/\lambda_{\parallel}$, that is, the ratio in inter-column shear to intra-column stretch moduli.  Assuming that resistance to intra-column stretching is much stronger than resistance to  inter-column sliding (which disrupts registry of the ``layered'', longitudinal order in the crystalline bundle), it is natural to expect that in general, $\chi_2^{-1} \ll1$.  The smallness of $\chi_2^{-1}$ is relevant because it sets a size scale $r_{sh} = \sqrt{\chi_2/|\chi_0|}$ at which the shear cost of twist dominates over the Frank elastic contributions to twist stiffness. Accordingly, the following discussion focusses on the cases where $r_{sh} \geq 1 \geq r_0$.

\subsubsection{Size-dependent twist}

\begin{figure}
\centering
\includegraphics[width=0.9\linewidth]{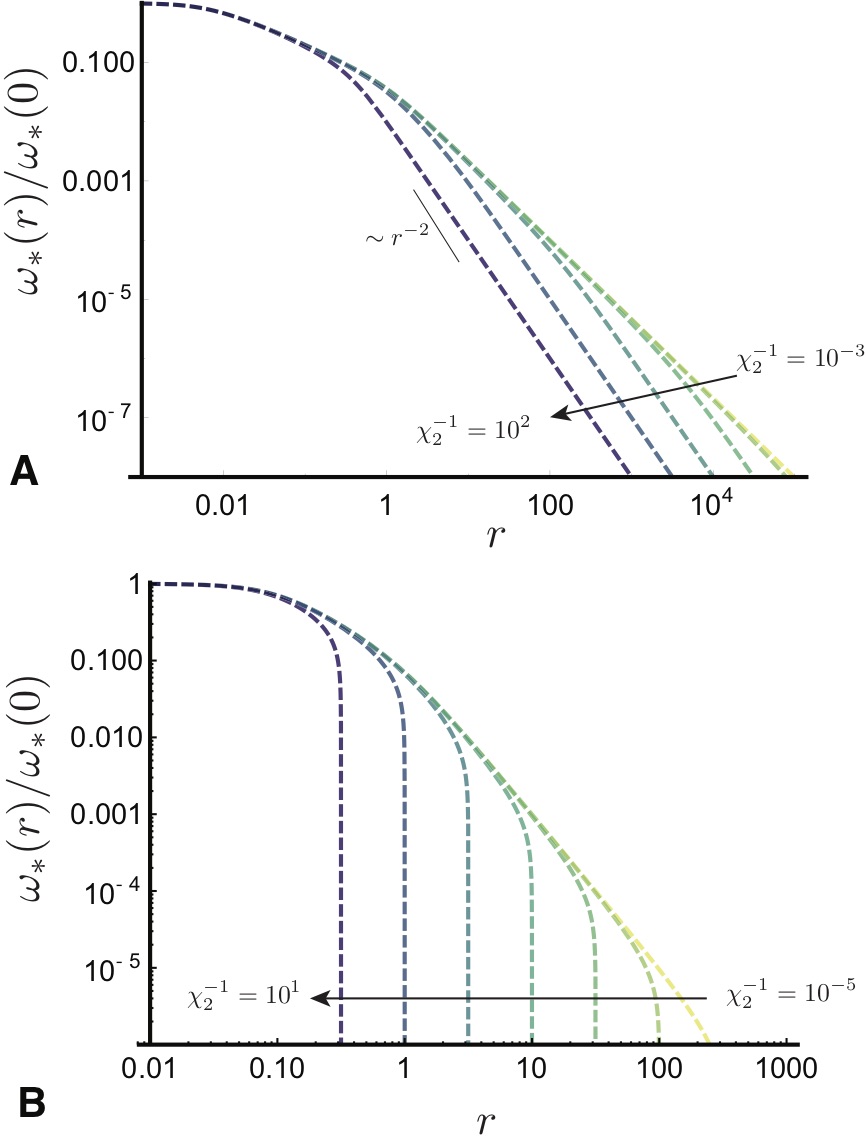}
\caption{Size-dependent twist for crystalline bundles of two varieties.  In (A), a chirality-driven, twist stable case with $Q=100$, $\chi_{0}^{-1}=1$ and for a series of reduced solid moduli, $\chi_2^{-1} =10^{-3} - 10^2$.  In (B), a spontaneously-twisting, achiral case with $Q=0$, $\chi_{0}^{-1}=-1$,  $r_0=0.1$ and for a series of longitudinal shear moduli, $\chi_2^{-1} =10^{-5} - 10^1$. }
\label{Figcryst_twist}
\end{figure}

Figure \ref{Figcryst_twist} shows plots of the equilibrium twist as function of size for two classes of crystalline bundles.  In Fig. \ref{Figcryst_twist}A, the dependence of $\omega_*$ on $r$ is plotted for chirality-driven, twist-stable bundles with fixed chirality ($Q=100$) and for a range of reduced solid moduil:  $\chi_2^{-1}=10^{-3} - 10^{2}$.   Comparing this behavior to the case of columnar bundles shown in Fig. \ref{chiral_col}A, non-zero shear modulus (i.e. $\chi_2^{-1}\neq 0$) leads to a more rapid untwisting of bundles with increased size.  Again, assuming $r_{sh} \geq 1$, the effect of longitudinal shear becomes dominant only well into the regime where columnar-strain drives untwisting.  Hence, crystalline bundles are characterized by a crossover from the columnar $\omega_* \sim r^{-4/3}$ regime, to the even more rapid $\omega_*\sim r^{-2}$ fall off deriving from the balance of chirality-driven and shear-elastic torques ($Q \approx \omega_* \chi_2^{-1} r^2$).  

For spontaneously-twisting, achiral bundles, shown in Figure Fig. \ref{Figcryst_twist}B, the effect of shear-elasticity of the crystalline phase is even
more profound.  This plot shows twist equilibria for a fixed value of $\chi_0^{-1} = -1$ and an increasing range of longitudinal shear rigidity, illustrating an abrupt (critical) transition from power law untwisting to the untwisted state at a critical bundle radius equal to $r_{sh}$.  This follows from the fact that  shear elasticity makes all crystalline bundles twist-stable at sufficiently large radii (i.e. $\chi^{-1} \geq 0$ for $r \geq r_{sh}$), and in the absence of intrinsic chirality, there is no mechanism to stabilize bundle twist when $\chi^{-1} >0$.

\subsubsection{Maximal self-limiting size}

\begin{figure}
\centering
\includegraphics[width=0.9\linewidth]{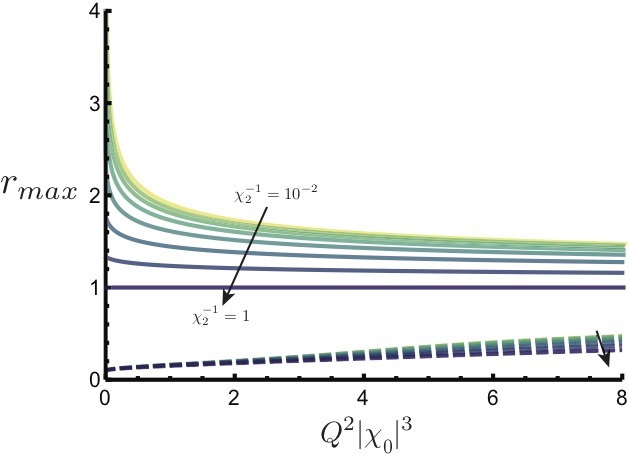}
\caption{Plot of maximal self-limiting bundle radius for crystalline bundles as a function of the combined chirality and twist-susceptibility, $Q^2 |\chi_0|^3$, and for a series of reduced solid moduli, $\chi_2^{-1}$.  The solid and dashed lines show the respective size of {\it twist-stable} ($\chi_0^{-1}>0$) and {\it twist-unstable} ($\chi_0^{-1}<0$) bundles.}
\label{Figcryst_rmax}
\end{figure}

Figure  \ref{Figcryst_twist} shows that the additional elastic cost of longitudinal shear of crystalline bundles leads to a reduced equilibrium twist in comparison to columnar bundles, and in particular, much more rapid rates of untwisting with increased size.  Figure \ref{Figcryst_rmax} shows the effect of the reduced twist of crystalline bundles on the range of their self-limiting size, as in Figure \ref{rmax_col}A, showing $r_{max}$ as a function of the combined parameters $Q^2 \chi_0^3$, but for series of increasing solid moduli:  $\chi_2^{-1}=10^{-2} - 10^{1}$.  

Generally speaking, the effect of non-zero $\chi_{2}^{-1}$ is to reduce the maximum size of stable self-limiting bundles, but its effect is most significant for the weak-chirality regime of twist-stable bundles (i.e. $\chi_{0}^{-1} >0$ and $Q \to 0$).  While the maximum size of columnar bundles is predicted to grow arbitrarily large as $Q\to0$, the maximum size of crystalline bundles never exceeds a length scale proportional to $r_{sh}$.  This is because the residual free energy from twist in the shear-dominated regime, $Q \omega_* \sim - r^{-2}$, cannot restrain the surface-energy drive (going as $r^{-1}$) towards bulk assembly.  Hence, for $\chi_{0}^{-1} >0$, stable finite bundles are restricted to the regime $r_{max} \lesssim r_{sh}$, which sets an upper limit to self-limitation under any chirality.   Notably, the effect on the maximum size {\it twist unstable bundles} (i.e. $\chi_0^{-1}<0$ ) deriving to crystalline shear elasticity is far more modest.  This is simply because, for the reasons described for the columnar case, such bundles lose thermodynamic stability well before reaching the size where crystalline shear elasticity becomes significant (i.e. $r_{max} < 1 < r_{sh}$).  

\section{Discussion and Conclusion}

\label{sec: discuss}

The previous sections describe a general theory for self-limitation induced by twist in self-assembled bundles and fibers adopting various degrees of internal order, from liquid crystalline (nematic) to crystalline (3D solid).  In addition to the comprehensive range of predictions for bundle thermodynamics (e.g. spectrum of power dependencies of twist on bundle size, and equilibrium size on surface energy), the analysis yields several broad conclusions.  

\begin{enumerate}

\item Intrinsic chirality, and corresponding preference for handed twist, is not essential for thermodynamically stable, finite bundles.  Spontaneous twist, which can arise in strictly {\it achiral} systems, introduces sufficient elastic penalties for lateral growth of bundles to limit their equilibrium diameter at low surface energy.  

\item Though they exhibit self-limitation (with and without chirality), the size-range of {\it twist-unstable} bundles is qualitatively distinct from {\it twist-stable} bundles.  Equilibrium {\it twist-unstable} bundles are limited to {\it microscopic} dimensions, comparable to few diameters in width, while {\it twist stable} bundles can extend well into to {\it mesoscopic} dimensions that far exceed the filament diameter.  In particular, the maximal size of twist-stable bundles becomes arbitrarily large (i.e. diverges) in the limit of vanishing chirality.  

\item Although {\it twist-unstable} bundles are generically limited to smaller sizes, they exhibit enhanced thermodynamic stability (i.e. retain finite diameter up to larger values of surface energy)  relative to their {\it twist stable} counterparts.

\end{enumerate}

The essential mechanism that underlies the distinctions between  twist-stable and twist-unstable bundles is the rate of bundle unwinding with increased diameter.  While the generalized elastic costs of bending and columnar and crystal strain resist twist equally in these distinct bundles, their driving forces for twist (torques) are not equal.  As shown schematically in Fig. ~\ref{introfig}, even as twist vanishes, chiral, twist-stable bundles are subject to a constant torque, while in the same limit, torques vanish in twist-unstable, achiral bundles.  The weakening of torques as bundles unwind towards $\omega \to 0$, accounts for the much more rapid size-dependent untwisting exhibited by twist-unstable bundles (see e.g. Figs.~\ref{chiral_col}A and ~\ref{achiral_col}A).  The more rapid untwisting of twist-unstable bundles, as characterized by a stronger power law decrease of twist with size, implies that the residual free energy gain of twisting is unable to restrain the drive to decrease surface energy by increasing to larger size.  Hence, {\it twist-unstable} bundles are driven to ``escape frustration''~\cite{hall2017} to the bulk state by untwisting at a smaller radius than {\it twist-stable} bundles which continue to resist untwisting by surface energy even at mesoscopic sizes. 

At present there is fairly little data from experimental studies on the inter-relation of structure and thermodynamics of twisted fibers, as many of physical parameters characterizing inter-filament assemblies remain difficult to measure, predict, or systematically vary, or otherwise, detailed structural measurements of intra-bundle order and morphology are limited by resolution limits of standard characterization techniques.  In an early study, Weisel and coworkers performed SEM measurements of the twist of purified and reconstituted fibrin bundles~\cite{weisel1987}.   While the bundles showed a range of finite diameters, from $\sim 30-50 {\rm nm}$, they maintained a constant helical pitch of roughly $200 {\rm nm}$.  This result is consistent with the equilibrium models above, assuming that bundle are assembling in the low surface energy regime and remain below their untwisting size.  Without data showing the rate of twist decrease with increased radius (presumably, as protofilament solubility is further decreased) it is not possible to definitely assess which of the mechanical costs of twists is responsible for limiting their lateral size.  

A more recent set of experiments has considered the possibility of self-twisting morphologies as a mechanism to regulate the thickness of fibers of marginally insoluble aggregates of methylcellulose (MC).  And while the detailed picture of the intra-fiber morphology of these aggregates remains a matter of some debate~\cite{ginzburg2016,li2017}, there is some evidence supporting the twisted fiber model of the assembly~\cite{morozova2018}.  Furthermore, the thermodynamics of twisted fiber assembly would seem to explain some salient experimental results.  Most notably, the thickness of self-assembled MC fibers is found to be independent of both molecular weight of MC chains and concentration, remaining within 17-19 nm over a wide range of conditions (corresponding to roughly $\sim 200$ chains in the cross section)~\cite{schmidt2018}.  This rather tight diameter regulation might support the interpretation that MC fibers are achiral, or at best, retain a weak chirality that may not propagate to twisted assembly.  Going beyond pure MC molecules, Morozova and coworkers ~\cite{morozova2017, morozova2018} modified the bending stiffness of the chains via grafting MC with a controlled density of oligomeric PEG side chains.  Upon assembly, it was found the fiber diameter {\it increased} with MC chain persistence length~\cite{morozova2018}.  Neglecting possible changes in other parameters (such as inter-chain cohesion), this increase was shown to be consistent with the predicted dependence of the maximal bundle size of an earlier theory~\cite{grason2007} for chirality driven twist in columnar bundles.  Assuming that the most significant change to MC chains upon grafting is increased bending stiffness, then one can go further to note that the only regime where bundle size increases with $K_3$, is indeed the limit of high-chirality, twist-stable bundles where equilibrium size grows as $R_* \sim K_3$ (and $R_{max} \simeq \Lambda_B \propto K_3^{1/2}$).  In other regimes of self-limiting assembly, either weak-chiralty, twist-stable, or achiral, twist-unstable, $R_*$ can be shown to {\it decrease} with $K_3$.  Hence, the experiments on PEG-modified MC fibers would then contradict the interpretation of MC assembly, as achiral or weakly-chiral, and suggest that the handed preference for twist in these aggregates is strong.  Additional experiments that, for example, could resolve the correlation of fiber twist with radius would be needed to clarify the role of chirality on MC fiber assembly.

In addition to the role of elastic parameters that characterize  resistance to bundle twist, this study highlights that at least two quantities are needed to describe the effective drive for inter-filament twist.  The first of which, the reduced chirality $Q$, is a dimensionless measure of the preferred inter-backbone twist.  Predicting the cholesteric pitch from the molecular structure and interactions of chiral molecules in dilute, liquid crystalline phases is notoriously challenging problem owing to the prominence of both positional and orientational fluctuations~\cite{straley1976, harris1999}.  In densely packed and oriented (at least nematic) bundles, many of these fluctuations are frozen out, and provided a sufficiently accurate model of chiral structure and interactions, predictions are available for the torques induced by chiral interactions between, helical biomolecules, such as DNA~\cite{kornyshev2007}.  Far less studied, at least from the perspective of inter-filamentary forces, is the twist stiffness, $\chi^{-1}$.  Steric considerations, for packing of inter-digitated disks, as in the columnar fibers of ~\cite{wales2009}, have been put forward to at least justify the {\it sign} of $\chi^{-1}$ and the appearance of spontaneous twist in achiral systems~\cite{nayani2015}.  On the other hand, modeling of pair-wise interactions between tubular filaments suggest that the both the sign and magnitude depend sensitive on the physical mechanism of inter-filament forces~\cite{cajamarca2014}.  For example, cohesive van der Waals interactions between tubular filaments generically lead to {\it twist-unstable} interactions, with a magnitude that varies with the ratio of interaction range to diameter.  In contrast, charged stabilized and osmotically-condensed tubular filaments are predicted to be {\it twist-stable}.  Taken together, these studies imply that twist stiffness exhibits a complex dependence on geometrical parameters of inter-filament/columnar packing as well as  competing mechanisms of inter-molecular forces at play in supramolecular systems.

Finally, I conclude by briefly noting two key physical effects that have not been considered in the generalized elastic theory presented here:  topological defects in the interior packing, and anisotropic surface shapes of bundles.  The former, dislocations and disclinations in the cross sectional lattice~\cite{bruss2012, bruss2013, grason2015} or tilt-grain boundaries in the ``smectic-like'' order of crystalline bundles~\cite{charvolin2014}, have been predicted to arise as means to mitigate the costs of geometric frustration associated with introducing twist to the respective 2D columnar and 3D solid order of bundles.  The latter effect of anisotropic cross-section shape may result in widely observed twisted, tape morphologies of bundles, and it has also been predicted to occur as elastically-driven response of surface shape to twist frustration~\cite{hall2016, hall2017}.   While these effects left out the present study, it is reasonable to expect that the would influence the quantitative, but not quantitative, conclusions presented above.  This is because these ``morphological mechanisms'' are capable of relaxing some, but not all of the frustration cost imposed by twist~\cite{hall2017}.  For example, incorporation of sufficiently many disclinations in twisted bundles can screen the power law growth of columnar strain with radius~\cite{grason2012, bruss2012}, but it does not eliminate the orientational (bending-induced) costs which also limit the lateral bundle radius.  Thus, the effect of forming twist-relaxing defects in bundle, for example, could potentially be captured by considering a defect-normalized values of the 2D elastic moduli, bundle assemblies that ``escape frustation'' of the 2D columnar lattice, would simply be described by the nematic limit of $\Lambda_B \to \infty$.

\section{Acknowledgements}
I would like to thank F. Bates, K. Dorfman, T. Lodge and D. Cleaver for stimulating discussions. I am grateful to D. Hall for detailed comments on this manuscript. This work was supported by the National Science Foundation under Grant DMR 1608862.  I would also like to acknowledge the hospitality of Aspen Center for Physics (NSF PHY 1607611) where much of this manuscript was completed.

\begin{appendix}

\section{Shear to stretching transition in crystalline bundles}

\label{crystalline}
The force balance equation for the longitudinal displacement $u_z$ for crystalline bundles, eq. (\ref{eq: fpar}), has the form 
\begin{equation}
\big( \lambda_\parallel \partial_z^2 + 2 \mu_\parallel \grad_\perp^2 \big) u_z = 0
 \end{equation}
and satisfies the boundary conditions,
 \begin{equation}
 \sigma_{zz}|_{z=\pm L/2} = 0;  \ \partial_r u_z|_{r=R}=0 ,
 \end{equation}
 where we have used $\grad \cdot \tv_\perp=0$ and $\hat{r} \cdot \tv_\perp=0$.  The solutions are harmonic functions and have the form,
 \begin{equation}
 u_z = \sum_{k} u(k) \sinh\big(\sqrt{2 \mu_\perp/\lambda_\perp} k z\big) J_0(k r) ,
 \end{equation}
 where vanishing radial stress at the sides of the bundle leads to the condition,
 \begin{equation}
 k_n R = x_n,
 \end{equation}
 where $x_n$ are the zeros $\partial_x J_0 (x)|_{x=x_n} = - J_1 (x_n)=0$.  The coefficients $u(k_n)$ derive from the cancelation of tension at the ends of the bundle $z=\pm L/2$,
\begin{multline}
u(k_n)= -\frac{(\Omega R)^2}{  \sqrt{2 \mu_\perp /\lambda_\perp} k_n \cosh \big( \sqrt{ \frac{2 \mu_\perp}{\lambda_\perp} } \frac{k_n L}{2} \big)} \\ \times \frac{2 J_2(k_n R)- k_n R J_3(k_n R)  }{  (k_n R)^2 J_0^2(k_n R) }   .
\end{multline}
From this, we have the longitudinal stress,
\begin{multline}
\sigma_{zz} = \lambda_\parallel (\Omega R)^2 \Big[ \frac{1}{2} \Big(-\frac{1}{2} +\frac{r^2}{R^2} \Big) \\ - \sum_{k_n >0} \frac{  \cosh\big(\sqrt{2 \mu_\perp/\lambda_\perp} k_n z\big) }{\cosh\big(\sqrt{2 \mu_\perp/\lambda_\perp} k_n L/2 \big)} \frac{2 J_2(k_n R)- k_n R J_3(k_n R)  }{  (k_n R)^2 J_0^2(k_n R) }   J_0(k_n r) \Big].
\end{multline}
Note that due to the term arising from the $k_n \to 0$ contribution to $\partial_z u_z$, the net stretching in the bundle vanishes at every $z$, i.e. $2 \pi \int _0^R dr r~\sigma_zz =0$.  Decomposing the shear stress in azimuthal and radial components we have,
\begin{equation}
\sigma_{\phi z} = \hat{\phi}_i \sigma_{i z} = \mu_{\parallel} \Omega r, 
\end{equation}
and 
\begin{multline}
\sigma_{r z} = \hat{r}_i \sigma_{i z} =  \sqrt{2 \mu_{\parallel} \lambda_\parallel} (\Omega R)^2  \\ \times \sum_{k_n >0}   \frac{  \sinh\big(\sqrt{2 \mu_\perp/\lambda_\perp} k_n z\big) }{\cosh\big(\sqrt{2 \mu_\perp/\lambda_\perp} k_n L/2 \big)} \frac{2 J_2(k_n R)- k_n R J_3(k_n R)  }{  (k_n R)^2 J_0^2(k_n R) }   J_1(k_n r)  .
\end{multline}
Example profiles are shown in Fig.~\ref{crystal}, where the the stretching vanishes within a zone of order $\sqrt{ \lambda_\perp/2 \mu_\perp} R$ from the ends of the bundle, and at its center, we find $\sigma_{zz} = \lambda_\parallel( \Omega^2/2) (r^2-1/2)$ consistent with $z$-independent stretching of outer filaments (which, in turn, loads the core filaments under compression). At the ends of the bundle, the relaxation of the tension, generates a zone of radial shear within the boundary zone, that decays to zero in the core of the bundle.  

From these results we can derive the elastic energy contributions for the twisted, 3D solid bundle.  First, the energy for longitudinal stretching,
\begin{multline}
\label{eq: stretch}
E_{st} \equiv \pi \lambda_{\parallel} \int_{-L/2}^{L/2} dz ~\int dr ~r u_{zz}^2 \\ = \frac{\lambda_{\parallel}}{96} V (\Omega R)^4 g_{st} \big(\sqrt{2 \mu_\parallel/\lambda_\parallel} L/R \big) ,
\end{multline}
where, $g_{st} (\alpha)$ is a dimensionless function characterizing the shear-to-stretch crossover,
\begin{multline}
g_{st} (\alpha) = 12 \sum_{n\geq 1} \Big[1- \frac{3 \sinh(\alpha x_n) -\alpha x_n }{ 2 \alpha x_n \cosh^2(\alpha x_n/2) } \Big] \frac{J_3^2(x_n)}{x_n^2 J_0^2(x_n)} \\ =  \left\{ \begin{array}{ll}\frac{5}{4}\alpha^3 & {\rm for} \ \alpha \ll 1 \\ \\ 1 & {\rm for} \ \alpha \gg 1 \end{array} \right. .
\end{multline}
The shear contributions break into radial and hoop components.  The radial shear contribution,
\begin{multline}
\label{eq: shear}
E_{sh,r} \equiv \pi (2 \mu_{\parallel}) \int_{-L/2}^{L/2} dz ~\int dr ~r u_{zr}^2 \\ = \frac{\lambda_{\parallel}}{96} V (\Omega R)^4 g_{sh} \big(\sqrt{2 \mu_\parallel/\lambda_\parallel} L/R \big) ,
\end{multline}
where, $g_{sh} (\alpha)$ is a non-monotonic function of $\alpha$ characterizing the build up of radial shear for small $L/R$, then the drop off to stretch dominated mechanics for $L/R \gg 1$,
\begin{multline}
g_{sh} (\alpha) = \sum_{n\geq 1} \Big[ \frac{\sinh(\alpha x_n) -\alpha x_n }{ \alpha x_n \cosh^2(\alpha x_n/2) } \Big] \frac{J_3^2(x_n)}{x_n^2 J_0^2(x_n)} \\ = \left\{ \begin{array}{ll}\frac{1}{3}\alpha^3 & {\rm for} \ \alpha \ll 1 \\ \\  \frac{c_1}{\alpha} & {\rm for} \ \alpha \gg 1 \end{array} \right. ,
\end{multline}
where $c_1 = 0.0616$. Finally, we have the shear of filaments separated along the hoop direction, which is independent of $R/L$ (as every concentric shell of filaments tilts by the same amount, $\Omega r$, relative to the central axis),
\begin{equation}
E_{sh,\phi}\equiv \pi (2 \mu_{\parallel}) \int_{-L/2}^{L/2} dz ~\int dr ~r u_{z \phi}^2 = \frac{\mu_\parallel}{8} V (\Omega R)^2 ,
\end{equation}
which is notably the ``Kirchoff beam" result of a twisted 3D solid rod.

\section{Equilibrium twist, radius and stability limit: general solution}

\label{generalsol}

Here, I summarize the equations of self-limiting bundle equilibrium, beginning with torque balance and equilibrium twist $\omega_*$, determined from the solution of,
\begin{equation}
\label{torque}
\omega \frac{\partial f}{ \partial \omega}= \beta \omega^4 + \chi^{-1} \omega^2 + Q \omega =0 ,
\end{equation}
where $\beta = r_0^2 + r^2+r^4$ and $\chi^{-1} = \chi_0^{-1}+ \chi_2^{-1} r^2$.  When $\chi^{-1} > \chi_*^{-1} = -3 \big( Q^2 \beta/4)^{1/3}$, there is one real solution, corresponding to the global minimum of the free energy,
\begin{equation}
\omega_* = \left\{ \begin{array}{r} - \sqrt{ \frac{4 \chi^{-1} }{3 \beta} } \sinh \bigg[ \sinh^{-1} \Big( \sqrt{ \frac{27 Q^2 }{ 4(\chi^{-1})^3} } \Big)/3  \bigg],  \  {\rm for \ } \chi^{-1} > 0 \\  \\ - {\rm sign}(Q) \sqrt{ \frac{-4 \chi^{-1} }{3 \beta}} \cosh \bigg[ \cosh^{-1} \Big( \sqrt{ \frac{27 Q^2 }{ 4(- \chi^{-1})^3}  } \Big)/3 \bigg], \\  {\rm for \ } 0> \chi^{-1} > \chi_*^{-1} \end{array} \right.
\end{equation}
When susceptibility is sufficiently negative, that is $\chi^{-1} < \chi_*^{-1} $, there are 3 real solutions, one maximum and 2 minima,
\begin{multline} 
\omega_*(n)=  - {\rm sign}(Q) \sqrt{ \frac{-4 \chi^{-1} }{3 \beta}} \cos \bigg[ \cos^{-1} \Big( \sqrt{ \frac{27 Q^2 }{ 4(- \chi^{-1})^3}  } \Big)/3  + \frac{2 \pi n}{3} \bigg],  \\  {\rm for \ } 0> \chi^{-1} > \chi_*^{-1} 
\end{multline}
where $n=0$, corresponds to the global minimum (for finite $Q$), $n=+1$ correspond to a local maximum and $n=+2$ corresponds to a metastable minimum.

Given these solutions, for equilibrium twist $ \omega_*(r)$, the  self-limiting bundle radius $r_*$ follows from minimization with respect to radius, 
\begin{equation}
\label{size}
r \frac{ \partial f }{\partial r} = \chi_{2}^{-1} r^2 \omega_*^2(r)+ (2r^2 + r^4)\omega_*^4(r)- \frac{\sigma}{r}
\end{equation}
which gives the equation of state relating the equilibrium finite size $r_*$ to the surface tension 
\begin{equation}
\sigma(r_*) = r_* \frac{ \omega^4_*(r_*)}{2} (r_*^2+2 r_*^4)+\chi_2^{-1}  \omega^2_*(r_*) r_*^2 .
\end{equation}

Finally, we can consider the limiting conditions for self-limiting assembly, namely the maximal finite size $r_{max}$ and minimal finite twist $\omega_{min}$, at which the bundles are in equilibrium with a surface energy $\sigma_{max}$, by the imposing condition such bundles are in equilibrium with bulk assembly (i.e. $f(\omega \to 0,r \to \infty) \to 0$), or,
\begin{equation}
\label{finitebulk}
\beta(r_{max}) \omega_{min}^4 + \chi^{-1} (r_{max}) \omega_{min}^2 + Q \omega_{min} + \frac{ \sigma_{max} }{r_{max} } = 0 .
\end{equation}
Combining this eq. (\ref{torque}) yields a parametric relationship between minimal finite twist and maximal size,
\begin{equation}
\omega_{min}^2 (r_{max}) = \frac{2 (\chi_0^{-1} - \chi_2^{-1} r_{max}^2 ) }{ ( r_{max}^4-r_{max}^2 -3 r_0^2)} . 
\end{equation}
Inserting this into eq. (\ref{size}) yields a parametric relation for maximal surface energy,
\begin{multline}
\sigma_{max} (r_{max})= 2 r_{max}^3  (\chi_0^{-1} - \chi_2^{-1} r_{max}^2 ) \\ \times \frac{ \big[ \chi_0^{-1} (1 + 2 r_{max}^2 ) - 2 \chi_2^{-1} (r_{max}^2 + 2 r_0^2) \big]}{(r_{max}^4-r^2_{max} - 3 r^2_0 )^2 } .
\end{multline} 
Inserting $\omega_{min} (r_{max}) $ into eq. (\ref{torque}) yields a parametric relation for reduced chirality at the stability limit
\begin{multline}
Q^2(r_{max}) = \frac{\chi_0^{-1} - \chi_2^{-1} r_{max}^2}{2 (r_{max}^4-r^2_{max} - 3 r^2_0 )^3} \\
\times \big[ \chi_0^{-1} (3 r_{max}^4 +r^2_{max}-r_0^2) - \chi_2^{-1} r_{max}^2 ( r_{max}^4 + 3 r_{max}^2 + 5 r_0^2) \big]^2 .
\end{multline}
Setting $\chi_2^{-1} = 0$ (for vanishing longitudinal shear modulus) yields the equations of state for the stability limit of finite columnar bundles, eqs. (\ref{eq: sigmax}) and (\ref{eq: chirality}).

\end{appendix}

%%%END OF MAIN TEXT%%%

%The \balance command can be used to balance the columns on the final page if desired. It should be placed anywhere within the first column of the last page.

%If notes are included in your references you can change the title from 'References' to 'Notes and references' using the following command:
%\renewcommand\refname{Notes and references}

%%%REFERENCES%%%
\bibliography{bib} %You need to replace "rsc" on this line with the name of your .bib file
\bibliographystyle{apsrev4-1} %the RSC's .bst file

\end{document}